\documentclass[useAMS,usenatbib,a4paper]{mn2e}

\usepackage{xspace}
\usepackage{graphicx}
\usepackage{amssymb}
\usepackage{amsmath}
\usepackage[pdf]{pstricks}

\newcommand{\apj}{ApJ}
\newcommand{\apjl}{ApJL}

\newcommand{\mnras}{MNRAS}

\newcommand{\etal}{et~al.~}
\def\spose#1{\hbox  to 0pt{#1\hss}}  
\newcommand{\lta}{\mathrel{\spose{\lower 3pt\hbox{$\sim$}}\raise  2.0pt\hbox{$<$}}}
\newcommand{\gta}{\mathrel{\spose{\lower  3pt\hbox{$\sim$}}\raise 2.0pt\hbox{$>$}}}

\newcommand{\be}{\begin{equation}}
\newcommand{\ee}{\end{equation}}

\newcommand{\deesq}{\, \mathrm{d}^2 \!}
\newcommand{\del}{\, \partial \!}

\newcommand{\bea}{\begin{eqnarray}}
\newcommand{\eea}{\end{eqnarray}}

\newcommand{\vect}[1]{\mathbf{#1}}
\newcommand{\operator}[1]{{#1}}

\newcommand{\kms}{\ifmmode  \,\rm km\,s^{-1} \else $\,\rm km\,s^{-1}  $ \fi }
\newcommand{\kpc}{\ifmmode  {\rm kpc}  \else ${\rm  kpc}$ \fi  }  
\newcommand{\pc}{\ifmmode  {\rm pc}  \else ${\rm pc}$ \fi  }  
\newcommand{\Msun}{\ifmmode {\rm M_{\odot}} \else ${\rm M_{\odot}}$ \fi} 
\newcommand{\Zsun}{\ifmmode {\rm Z_{\odot}} \else ${\rm Z_{\odot}}$ \fi} 
\newcommand{\yr}{\ifmmode yr^{-1} \else $yr^{-1}$ \fi} 
\newcommand{\hMsun}{\ifmmode h^{-1}\,\rm M_{\odot} \else $h^{-1}\,\rm M_{\odot}$ \fi}

\def\zd{z_{\rm d}}
\def\zs{z_{\rm s}}

\newcommand{\comment}[1]{}
\newcommand{\comments}[1]{}

\newcommand{\new}[1]{{#1}}

\newcommand{\al}{\vect{\alpha}_l}
\newcommand{\ap}{\vect{\alpha}_p}
\renewcommand{\b}[1]{\beta_{#1}}

\def\ioa{Institute of Astronomy, University of Cambridge,
  Madingley Rd, Cambridge, CB3 0HA, UK}

\def\collettemail{\tt t.collett@ast.cam.ac.uk}

\title[Cosmological Constraints from the double source plane lens SDSSJ0946+1006]
{Cosmological Constraints from the double source plane lens SDSSJ0946+1006}

\author[Collett \etal]{%
  Thomas~E.~Collett\thanks{\collettemail},
  Matthew~W.~Auger
  \\\ioa
}

\begin{document}
             
\date{to be submitted to MNRAS}
\pagerange{\pageref{firstpage}--\pageref{lastpage}}\pubyear{2014}

\maketitle           

\label{firstpage}


\begin{abstract} 

We present constraints on the equation of state of dark energy, $w$, and the total matter density, $\Omega_{\mathrm{M}}$, derived from the double-source-plane strong lens SDSSJ0946+1006, the first cosmological measurement with a galaxy-scale double-source-plane lens. By modelling the primary lens with an elliptical power-law mass distribution, and including perturbative lensing by the first source, we are able to constrain the cosmological scaling factor in this system to be $\beta^{-1}=1.404 \pm 0.016$, which implies $\Omega_{\mathrm{M}}= 0.33_{-0.26}^{+0.33}$ for a flat $\Lambda$CDM cosmology. Combining with a CMB prior from Planck, we find $w$ = $-1.17^{+0.20}_{-0.21}$ assuming a flat $w$CDM cosmology. This inference shifts the posterior by $1\sigma$ and improves the precision by 30 per cent with respect to Planck alone, and demonstrates the utility of combining simple, galaxy-scale multiple-source-plane lenses with other cosmological probes to improve precision and test for residual systematic biases.

\end{abstract}

\begin{keywords}
cosmological parameters --  gravitational lensing

\end{keywords}

\setcounter{footnote}{1}


\section{Introduction}
\label{sec:intro}

The current concordance cosmology of $\Lambda$CDM gives a remarkably good fit to current observational data \citep{planck16,wmap9,percival}, but a tension seems to be emerging between different cosmological probes. For example, assuming flat $\Lambda$CDM the recent Planck constraints on the Hubble constant \citep{planck16} are significantly lower than those found by local measurements using supernovae \citep[][but see \citealt{efstathiou}]{freedman2012,riess2011} and strong lens time delays \citep{suyu2013}. Although this discrepancy may simply be due to the presence of unknown systematic errors, it might also signal physics beyond the flat $\Lambda$CDM model; \textit{independent} cosmological probes are therefore needed to test the assumption that the universe is spatially flat and that the dark energy is a cosmological constant.

\begin{figure}
\includegraphics[width=\columnwidth,clip=True]{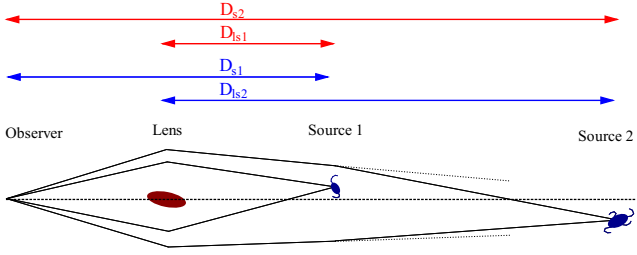}
\caption {Sketch of a double source plane lens system. The cosmological scaling factor $\beta$ is the product of $D_{\text{ls}1}$ and $D_{\text{s}2}$ (both in red) divided by the product of $D_{\text{ls}2}$ and $D_{\text{s}1}$ (both in blue). For a singular isothermal sphere, where the first source has no mass, $\beta$ is the ratio of Einstein radii. Figure taken from \citet{collett2012}.}
\label{fig:bench}
\end{figure}

Strong gravitational lensing is potentially a powerful tool to test cosmological models \citep{witt, kochanek, saha, schechter, oguri2007, oguri2012, suyu2010, suyu2013, gavazzi, collett2012}, due to its sensitivity on the distances between components of the lens system (e.g., the observer, the foreground massive lensing object, and any background lensed sources). In principle, measurements of the Einstein radius and enclosed lens mass are sufficient to constrain cosmological parameters \citep{grillo2008, biesiada2010}, but robustly inferring the lensing mass is degenerate with the choice of lens density profile. To make robust inference on cosmological parameters -- without making strong assumptions about the lens mass distribution -- additional information is required. Gravitational lens systems with two background sources at different redshifts (schematic shown in Figure~\ref{fig:bench}) provide sufficient information to make precision measurements of cosmology.

In \citet{collett2012} we showed that double source plane lenses (DSPLs) can be a useful, complementary cosmological probe, allowing the dark energy equation of state to be constrained independently of the Hubble constant. \citet{jullo2010} constrained cosmological parameters using 12 multiply-lensed sources behind the cluster Abell 1689, but the sparsity of lensed images and the clumpy mass distribution in clusters makes the measurement difficult; the systematic uncertainties are likely to be large \citep{zieser2012} and much of the information provided by the multiple background sources may need to be used to infer the complexity of the lensing mass distribution. Galaxy-galaxy strong lenses, on the other hand, tend to be very well fitted with simple mass distributions (e.g., Vegetti et al., submitted) and therefore may be preferable objects for testing cosmology.

In this work we present strong lensing models of SDSSJ0946+1006 (hereafter J0946), a DSPL serendipitously discovered by \citet{gavazzi}, and we use these lens models to estimate the cosmology-dependent scale factor $\beta$ that governs the relative lens strength acting on each of the two source planes. In Section \ref{sec:theory} we outline the relevant theory of multiple source plane lensing. In Section \ref{sec:modelling} we develop a framework for fitting multiple source plane lenses with regularized pixellated sources and apply this technique to J0946. In Section \ref{sec:cosmology} we convert our measurement of $\beta$ into constraints on the cosmological parameters $w$ and $\Omega_{\mathrm{M}}$. We discuss and conclude in Section \ref{sec:discussion}.

\section{Cosmography with double source plane strong lenses}
\label{sec:theory}

\new{
In this section we briefly review the theory of compound lensing, following \citet{SEF} and summarize how observations of multiple-source-plane strong lenses can be used to constrain cosmology. 
For a single-source-plane lens, the lens equation can be written as
\begin{equation}
\label{eq:SSPL}
\mathbf{y} = \mathbf{x} - \mathbf{\alpha}(\mathbf{x}),
\end{equation}
where $\mathbf{y}$ is the position of the source on the source plane and  $\mathbf{x}$ is the position in the image plane. $\mathbf{\alpha}(\mathbf{x})$ is the scaled deflection caused by the lens, given by 
\begin{equation}
{\alpha}(\mathbf{x})~=~\frac{1}{\pi} \int \deesq x' \kappa(\mathbf{x'})\frac{\mathbf{x}-\mathbf{x'}}{|\mathbf{x}-\mathbf{x'}|^2}
\end{equation}
where $\kappa(\mathbf{x})$ is the lens surface mass density, $\Sigma(D_l \mathbf{x})$, scaled by the lensing critical density, $\Sigma_{\mathrm{crit}}$,
\begin{equation}
\kappa= \frac {\Sigma}{\Sigma_{\rm cr}(\zd,\zs)}
\end{equation}
with,
\begin{equation}
\label{eq:sigmacrit}
\Sigma_{\rm cr}(\zd,\zs) \equiv \frac{c^2 D_{\rm s}}{4 \pi G D_{\rm l} D_{\rm ls}}
\end{equation}
and $D_{\mathrm{ij}}$ are angular diameter distances between observer, lens and source. 
Changing the source redshift alters only the angular diameter distances in Equation \ref{eq:sigmacrit}, whilst the other terms in Equations \ref{eq:SSPL} to \ref{eq:sigmacrit} are unchanged. Hence for two photons passing through the same point in the lens plane, but originating on different source planes, the ratio of scaled deflection angles is given by the cosmological scaling factor, $\beta$,
\begin{equation}
\label{eq:beta}
\frac{\mathbf{\alpha_1}}{\mathbf{\alpha_2}} = \frac{D_{ls1} D_{s2}}{D_{s1} D_{ls2}} \equiv \beta.
\end{equation}
In the special case of a singular isothermal sphere lens, $\beta$ is simply the ratio of Einstein radii.

When multiple sources are present, the mass of the first source will lens the light coming from the second. Including this effect requires the use of the full multiple-lens-plane formalism, as given in \citet{SEF}. For a system with $j-1$ lens planes (any of which can be masses and/or sources), and a final source, the position on each plane can be calculated by ray-tracing back iteratively, using 
\be
\label{eq:multilensequation}
\vect{x}_j=\vect{x}_1 - \sum_{i=1}^{j-1}\beta_{ij}\vect{\alpha}_{i}(\vect{x}_i)
\ee
where the reduced deflections $\vect{\alpha}_{i}$ are the physical deflections rescaled for the final source plane
\begin{equation}
\vect{\alpha}_i = \frac{D_{is}}{D_{s}}\hat{\vect{\alpha}}_i
\end{equation}
and $\beta_{ij}$ is the cosmological scaling factor given in Equation \ref{eq:beta} (from hereon we use $\beta$ as shorthand for $\beta_{12}$). In the case of a double-source-plane lens, the lens equation for photons originating on the first and second source planes are respectively given by
\label{eq:DSPLs1}
\begin{equation}
\mathbf{y}^{\rm {S1}} = \mathbf{x} - \beta \mathbf{\alpha_{\rm {l}}}(\mathbf{x}),
\end{equation}
and
\begin{equation}
\label{eq:DSPLs2}
\mathbf{y}^{\rm {S2}} = \mathbf{x} - \mathbf{\alpha_{\rm {l}}}(\mathbf{x}) - \mathbf{\alpha_{\rm {S1}}}(\mathbf{x}- \beta \mathbf{\alpha_{\rm {l}}}(\mathbf{x})).
\end{equation}

Angular diameter distances, and hence $\beta$, are functions of redshift and the cosmological parameters,

\begin{equation}
D_{\text{ij}}~=~ {c/H_0 \over {(1+z_{\text{j}})}}\left( {\mathrm{sinn}\!\left( \sqrt{|\Omega_k|} \int_{z_{\text{i}}}^{z_{\text{j}}} {\mathrm{d} z \over E(z)}\right)\over \sqrt{|\Omega_k|}}\right)
\label{Da}
\end{equation}
where $\mathrm{sinn}(x)~=~\sin(x)$, $x$, or $\sinh(x)$ for open, flat, or closed universes respectively, and $E(z) \equiv {H(z) \over H_0}$ is the normalised Hubble parameter. For a $w$CDM cosmology,
\begin{equation}
E^{w\text{CDM}}~=~\!\sqrt{{\Omega_{\mathrm{M}}(1\!+\!z)^{3}\!+\!\Omega_k(1\!+\!z)^2\!+\!(\Omega_{\mathrm{de}})(1\!+\!z)^{3(1+w)}}}.\label{Ez}
\end{equation}
Equation \ref{Ez} holds if the dark energy equation of state, $w$, is constant; fixing $w=-1$ and $\Omega_{k}=0$ gives the concordance $\Lambda$CDM cosmology. Since angular diameter distances are inversely proportional to $H_0$, the ratio $\beta$ is a function only of $w$, $\Omega_\text{M}$, $\Omega_\text{k}$ and the redshifts of the lens and sources.
}

\section{Modelling the double source plane system SDSSJ0946+1006}
\label{sec:modelling}

J0946 was discovered by \citet{gavazzi} as part of the Sloan Lens ACS Survey \citep{bolton06,bolton08}. The lens is a massive early-type galaxy at $z_l = 0.222$ and the first source has a redshift of $z_{s1} = 0.609$ \citep{gavazzi} while the second source has a photometric redshift of $z_{s2} \approx 2.4$ \citep{sonnenfeld}. \citet{gavazzi} reported that for J0946 the Einstein radii are $1.43\pm0.01$ and $2.07\pm0.02$ arcseconds for the inner and outer rings, hence $\beta^{-1}$ is approximately 1.45. Subsequently, \citet{sonnenfeld} have investigated the system in more detail, explicitly including the effects of the lower-redshift source and using a two-component lensing model to constrain the dark and stellar mass distributions.

\begin{figure}
\includegraphics[width=0.95\columnwidth,clip=True]{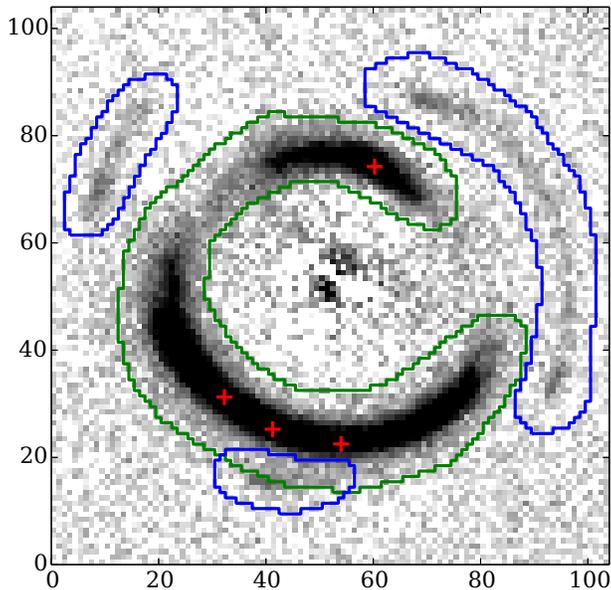}

\caption{The galaxy-subtracted $HST$ F814W image of SDSSJ0946+1006. We model only the data contained within the union of the green and blue masks. The first source is modelled as only contributing to the flux observed within the green mask, and the second source is assumed to only have non-zero flux within the blue mask. The red crosses mark the 4 pixels which we map back on to the first lens plane and use to determine the centroid of the first source's mass.}
\label{fig:masks}
\end{figure}

Here we fit a new lens model to simultaneously constrain the mass in the foreground lens, the velocity dispersion of the lower-redshift source, and the cosmology-dependent term $\beta$. We use $HST$ ACS imaging data in the F814W filter that have been drizzled to a $0\farcs05$ pixel scale and our point spread function model is taken from a bright, unsaturated star in the image. We first subtract the lensing galaxy by simultaneously fitting the foreground galaxy and both background sources in the same framework as described in \citet{auger11}. Briefly, we assume that the lensing mass is a singular isothermal ellipsoid with external shear, and both sources are modelled with Sersic profiles. The lensing galaxy has a complicated photometric profile and shows evidence for interactions \citep[also see][]{sonnenfeld}, but we find that it is well-fit using three Sersic components. The goal of this modelling is to robustly characterise and remove the light from the lensing galaxy, and the result is shown in Figure \ref{fig:masks}.

\subsection{Lens Model}
We fit the lensed images shown in Figure \ref{fig:masks} with two pixellated sources lensed by an elliptical power-law mass distribution with external shear.  We apply 2 overlapping masks (shown in Figure \ref{fig:masks}) where we allow each source to place flux, leaving us a data vector of the $N_D$ pixels within the union of the two masks, that we fit with two sources ($s_1$ and $s_2$) of $N_{s1}$ and $N_{s2}$ pixels respectively. 

Our primary lens model has 6 free parameters, two for position ($x,y$), two for ellipticity ($q,\theta_q$), the power-law index ($\eta$) and the Einstein radius ($b_l$, the characteristic scale of deflections by the lens for the first source plane), plus 2 further parameters for external shear (the shear magnitude and position angle, $\gamma_{\mathrm{ext}}$ and $\theta_\gamma$, both defined in terms of their effect on the first source). We model the mass of the first source as an isothermal sphere with one free parameter, the Einstein radius ($b_{s1}$). In our model, the centroid of this mass is a deterministic function of the primary lens model. Experience tells us that mass and light are closely aligned, but we do not know the unlensed light distribution for the first source a-priori. To retain the prior that the mass and light are co-spatial, we map 4 of the brightest first source pixels in the image plane back onto the source plane according to the current primary lens model (see Figure \ref{fig:masks}) and take the mean of these 4 positions as the centroid of the source mass. Because lensing conserves surface brightness, bright image plane pixels should always map onto bright source plane pixels, but in practice our inference is insensitive to our choice of centroid as long as the mass is placed approximately near the light. The final free parameter in the model is the \new{inverse of the} cosmological scaling factor $\beta$, as defined in Equation \ref{eq:beta}. In total our model has 10 free non-linear parameters, and we assume uninformative uniform priors for each of these.

\subsection{Modelling DSPLs with pixellated sources}

\begin{figure*}
\begin{center}
\includegraphics[trim=.001cm 0cm 0cm 0cm, clip=True, width=0.95\linewidth]{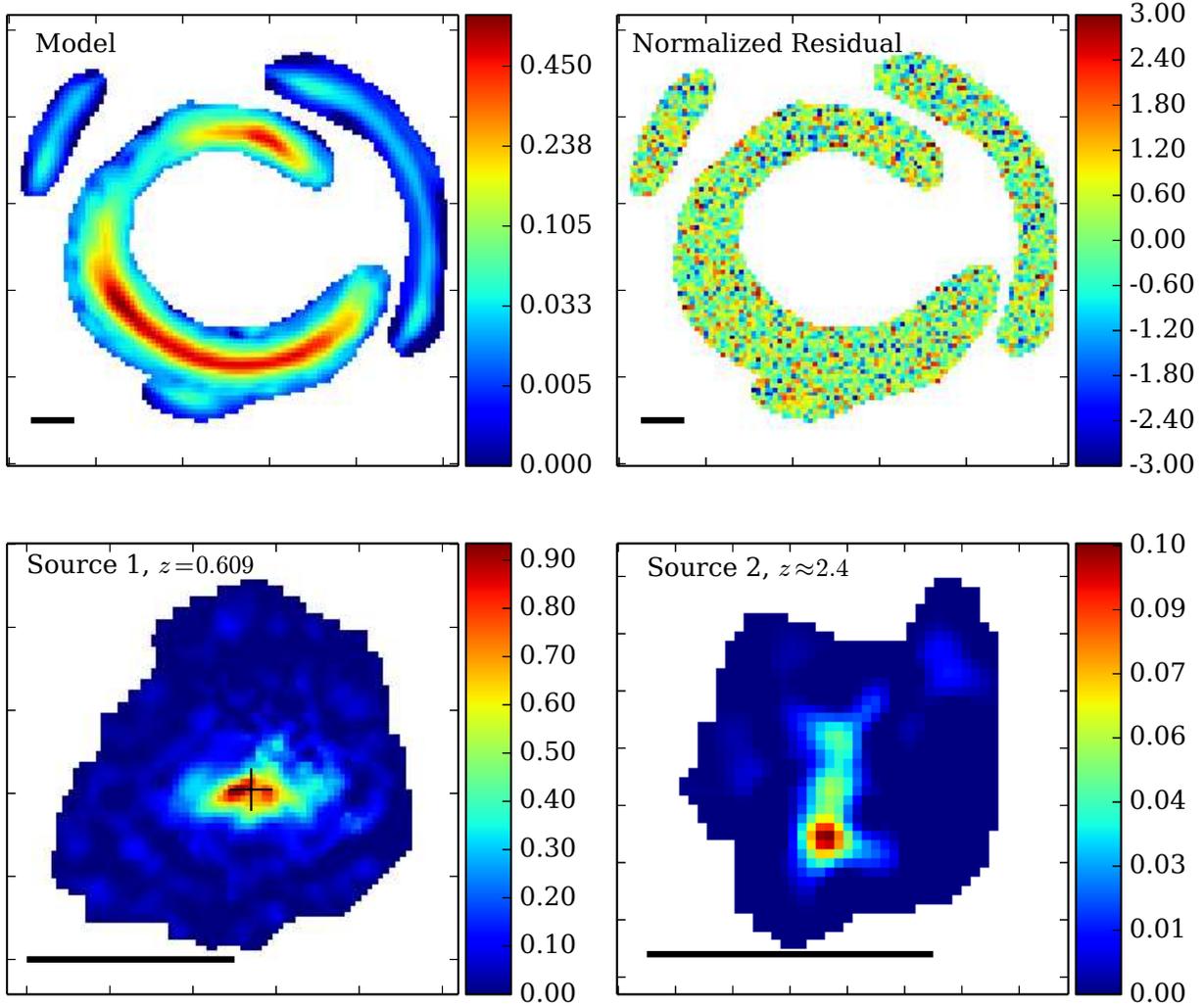}
\end{center}
\caption{From top left clockwise. 1) Most probable model for the F814W image of SDSSJ0946+1006, the colour scale is non-linear. 2) Normalized residual, (image-model)/$\sigma$, where $\sigma$ is the uncertainty in each pixel. 3) Most probable first source, the centroid of this source's mass is shown by the black cross. 4) Most probable second source. In all images the black bar shows a 0.5 arcsecond scale.
}\label{fig:bestmodel}
\end{figure*}

\begin{figure*}
\begin{center}
\includegraphics[trim=.001cm 0cm 0cm 0cm, clip=true, width=0.95\linewidth]{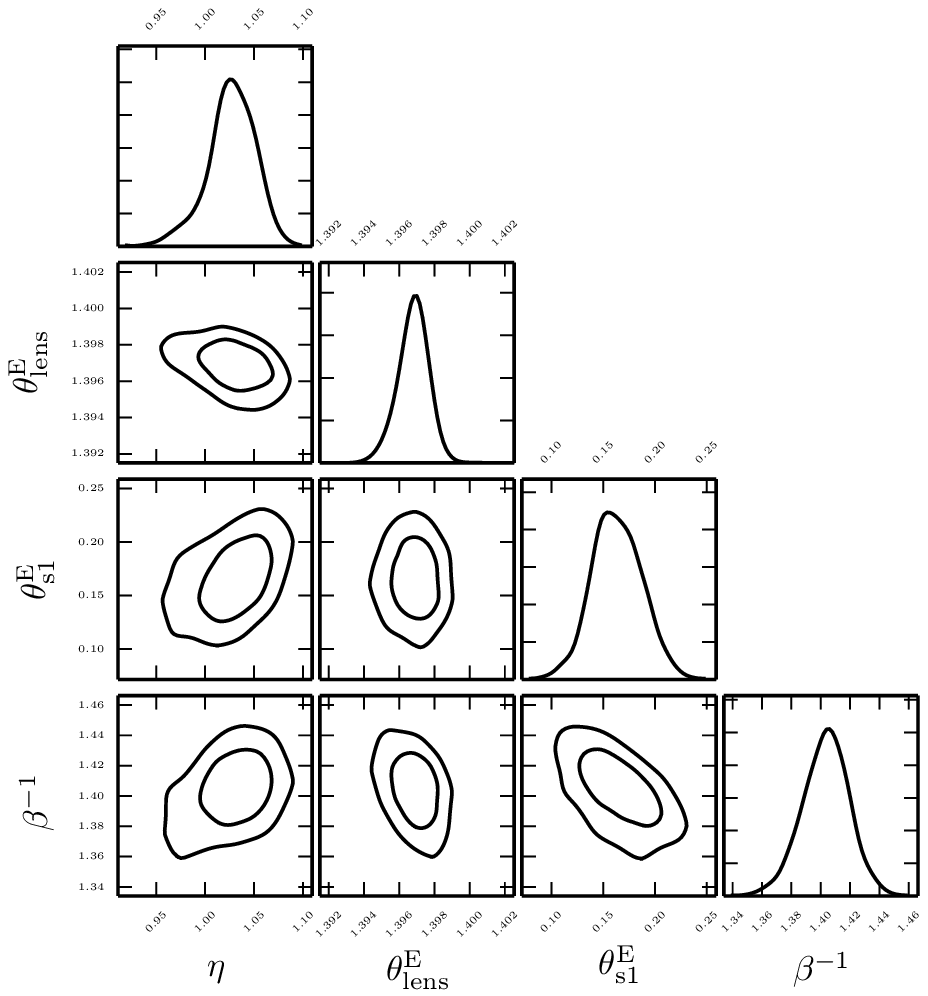}
\end{center}
\caption{The posterior for $\beta^{-1}$ from our model, and the 3 model parameters that are most degenerate with it: the logarithmic slope of the projected density for the first lens ($\eta=1$ corresponds to isothermal), and the Einstein radii of the primary lens and the first source ($\theta_{\mathrm{lens}}^{\mathrm{E}}$ and $\theta_{\mathrm{s1}}^{\mathrm{E}}$ respectively) in arcseconds. The contours show the 68 and 95\% confidence regions.}
\label{fig:bestpars}
\end{figure*}

Our modelling follows the semi-linear approach of \citet{warren+dye}, where we linearly solve for the optimal pixellated source during each iteration of non-linearly sampling over the mass model parameters.

\subsubsection{Modelling multiple pixellated sources with a fixed lens model}

To solve for the source and lens parameters we first pick a lens model and reverse ray-trace the image plane pixels back onto each source plane using the elliptical power-law deflection solver of \citet{barkana} and the multiple lens-plane equation (Equation \ref{eq:multilensequation}). For each source we then define a square grid that is sufficiently large to encompass the de-lensed position of each image pixel contained within the corresponding mask\footnote{This partially adaptive grid is designed so that choosing a fixed number of source grid pixels, i.e., $N_{s1}$ and $N_{s2}$, does not bias the fit towards any particular lens model. For a non-adaptive source grid, models with high magnification would fill a smaller region of the source plane (and hence have fewer degrees of freedom) than those with low magnification}. With the source grids now defined, and the de-lensed position of each image pixel now known, we can apply the bilinear interpolation method of \citet{treu+koopmans} to generate the lensing matrix, $\operator{l}$.

For a fixed lens model, we can write down the equation that relates the data vector (i.e., the observed pixels in the image plane) to the two sources:
\begin{equation}
\vect{d}=\operator{f}\vect{s} +\vect{n}
\end{equation}
where $\vect{d}$ is the data vector, $\vect{s}$ is the concatenated source vector ($s_1$,$s_2$), and $\vect{n}$ is the noise vector characterized by the covariance matrix, $C_{\text{D}}$. $\operator{f}$ is the ($N_{d}$~by~$N_{s1}$+$N_{s2}$) matrix that maps source flux onto the image plane and is the product of the point spread function operator ($\operator{p}$) and the lensing operator: $\operator{f} = \operator{p}\operator{l}$.

Without any prior on the smoothness of the source (regularization), the merit function for this model is \citep{warren+dye}
\be
{\exp\left(-E_D\right)}
\ee
where
\be
E_D={\frac{1}{2} \left(\operator{f}\vect{s} - \vect{d}  \right)^{\mathrm{T}} {C_{\text{D}}}^{-1} \left(\operator{f}\vect{s} - \vect{d} \right)}.
\ee
\citet{warren+dye} and \citet{suyu2006} show that the most likely source is thus given by
\be 
\label{eq:sML}
\mathbfit{s}_{\mathrm{ML}} = \operator{F}^{-1} \mathbfit{D},
\ee
where
\be
\label{eq:F}
\operator{F} = \operator{f}^{\mathrm{T}} {C_{\text{D}}}^{-1} \operator{f}
\ee
and
\be
\label{eq:D}
\mathbfit{D}= \operator{f}^{\mathrm{T}} {C_{\text{D}}}^{-1} \vect{d}.
\ee

However, this may give sources that do not look like reasonable astrophysical objects, and when the number of source pixels is large there may be unreasonably many degrees of freedom. This problem can be overcome by the introduction of a regularization term \citep{wallington, warren+dye} that prefers the source to be smooth. This regularization adds additional terms to the merit function that penalize spiky sources, such that the full merit function is given by
\be
\label{eq:merit}
{\exp\left(-E_D - \lambda_1 E_{s1} - \lambda_2 E_{s2} \right)}
\ee
where
\be
E_{si}={\frac{1}{2} \left(\vect{s_i}^{\mathrm{T}} \operator{R_i} \vect{s_i}\right)}
\ee
and $R_i$ is the regularization matrix. This regularization modifies Equation \ref{eq:sML} to
\be 
\label{eq:sMP}
\mathbfit{s}_{\mathrm{MP}} = \operator{A}^{-1} \mathbfit{D},
\ee
where the matrix \operator{A} is defined as 
\be
\operator{A}=\operator{F} +\begin{pmatrix} \lambda_1 \operator{R_{s1}} & 0 \\ 0 & \lambda_2 \operator{R_{s2}} \end{pmatrix}.
\ee

\citet{suyu2006} showed how to optimize the strength of this regularization, $\lambda$, for a single source. Under certain assumptions \citep{suyu2006}, the optimal value of $\lambda$ is found by solving the equation 
\be
\label{eq:optimallambda}
\frac{\mathrm{d}}{\mathrm{d}\log\lambda} \log P(\vect{d}|\lambda,\operator{f},r)=0
\ee
where $r$ represents the choice of regularization.

When there are multiple sources, Equation \ref{eq:optimallambda} can be generalized to find the optimal regularization for each,
\be
\frac{\partial}{\partial\log\lambda_i} \log P(\vect{d}|\{\lambda\},\operator{f},r)=0.
\ee
Generalizing Equation 19 of \citet{suyu2006} to include 2 sources, it can be shown that
\bea
\label{eq:EvidFull}
\log P(\vect{d}|\{\lambda\},\operator{f},r) \!&\!\!\!=&\!\!\!\!- \lambda_1 E_{\mathrm{S1}} -\lambda_2 E_{\mathrm{S2}} - E_{\mathrm{D}}-  \frac{1}{2}\log(\det\operator{A})
\nonumber \\ & &\!\!\!\!+ \frac{N_{\rm s1}}{2}\log\lambda_1 + \frac{N_{\rm s2}}{2}\log\lambda_2
\nonumber \\ & &\!\!\!\!+  \text{ terms independent of } \{\lambda_1,\lambda_2\}
\eea
Hence the optimal value for each regularization constant is given by
\be
2\lambda_i E_{si} = N_{si}-\lambda_i  \frac{\partial}{\partial  \lambda_i}(\log (\det \operator{A}))
\ee
which must be solved iteratively since both $\operator{A}$ and $E_{si}$ are functions of $\lambda_1$ and $\lambda_2$\footnote{$\frac{\partial}{\partial  \lambda_i}(\log (\det \operator{A}))$, can be efficiently solved numerically using the Cholesky root of $\operator{A}$; $\log (\det \operator{A}) = 2 \sum_j{\log[({A^{1/2}})_{jj}] }$}.

By picking a specific coarseness for the source pixel grid, we are implicitly assuming the source to be smooth on scales smaller than the pixel size. By using a fine pixel grid, we can minimise this assumption and allow the Bayesian evidence to choose the regularization rather than an ad-hoc choice of pixel size. To balance the need for a fine grid and computational demands, we use source grids that have 80 pixels on a side for the first source and 50 pixels on a side for the second source. Fewer pixels are needed for the second source since the signal to noise of the first source's arc is much greater than the second's, hence the optimal regularization for the second source is likely to be significantly stronger than for the first source. Our results do not change significantly if the number of source pixels is changed.

\subsubsection{Determining the mass model}

Whilst solving for the source is a linear problem, we determine the lens model parameters by using the parallel tempered ensemble sampler of \citet{emcee} to sample this 10 dimensional space, solving Equation \ref{eq:sMP} at each iteration in order to evaluate the merit function of Equation \ref{eq:merit}. By using the parallel tempered ensemble sampler, we are able to have high confidence that we are sampling the full posterior and not a single island of high likelihood. Ideally we would optimize the regularization for each iteration of the sampler, however this is computationally expensive; we find \citep[as also found by][]{vegetti} that it is sufficient to optimize the two source regularizations at the most probable lens model\footnote{The most probable lens model can be solved iteratively, positing large regularizations, optimizing the lens model then solving for the optimal regularizations and repeating. One iteration is typically sufficient \citep{vegetti}.} and fix these values throughout the chain.

\subsection{Modelling results}
\label{sec:modellingresults}

\begin{table*}
\caption{The median and 68\% confidence bounds on the 10 marginalized parameters of our lens model.}
\begin{tabular}{@{}cccccccccc}
$\!\!x_{\mathrm{lens}}$&$y_{\mathrm{lens}}$&$q_{\mathrm{lens}}$&$\theta_q$&$\eta_{\mathrm{lens}}$&$\theta^{\mathrm{E}}_{\mathrm{lens}}$&$\theta^{\mathrm{E}}_{\mathrm{s1}}$&$\gamma_{\mathrm{ext}}$&$\theta_\gamma$&$\beta^{-1}$\\

$51.885^{+0.057}_{-0.038}\!\!$&$51.429^{+0.040}_{-0.036}\!\!$&$0.946^{+0.009}_{-0.005}\!\!$&$30.6^{+4.0}_{-4.8}\!\!$&$1.027^{+0.023}_{-0.025}\!\!$&$1.397^{+0.001}_{-0.001}\!\!$&
$0.161^{+0.025}_{-0.021}\!\!$&$0.069^{+0.002}_{-0.003}\!\!$&$-27.20^{+0.75}_{-0.56}\!\!$&$1.405^{+0.014}_{-0.016}\!\!$

\label{tab:modelresults}
\end{tabular}
\end{table*}

Applying the techniques developed in Section \ref{sec:modelling}, the marginalized parameter constraints are given in Table \ref{tab:modelresults} and we show 2D projected sample distributions for $\eta$, $\theta_{\mathrm{lens}}^{\mathrm{E}}$, $\theta_{\mathrm{s1}}^{\mathrm{E}}$ and $\beta^{-1}$ in Figure \ref{fig:bestpars}; the results of our modelling can be summarised as follows:
\begin{itemize}
\item{The mass distribution of J0946 in the region probed by the two Einstein rings is very close to that of an isothermal sphere with a projected logarithmic density slope of $\eta$ = $1.027^{+0.023}_{-0.025}$ (An isothermal sphere has $\eta$ = 1)}
\item{The Einstein radius of the primary lens is very precisely determined. The primary lens has an Einstein radius of $\theta_\mathrm{lens}^{\mathrm{E}}$ = $1.397^{+0.001}_{-0.001}$, similar to the values found by previous models of J0946 \citep{bolton08,gavazzi,vegetti,sonnenfeld}\footnote{We note that here the Einstein radius $\theta_\mathrm{lens}^{\mathrm{E}}$ is the circularised (i.e., intermediate axis) radius within which the mean surface mass density is equal to the lensing critical density (i.e., the mean convergence is unity). Other authors may report Einstein radii measured along the major axis of the elliptical mass distribution or defined as a lens strength in some other way, but when we apply this definition they are all in close agreement}.}
\item{By modelling the full arcs we have been able to make precise inference on the Einstein radius of the first source, and we find that $\theta_\mathrm{s1}^{\mathrm{E}}$ = $0.161^{+0.025}_{-0.021}$, strongly excluding the zero mass case. This implies a velocity dispersion of $\sim97\pm7~{\rm km~s^{-1}}$.}
\item{The cosmological scaling factor $\beta^{-1}$ is $1.405^{+0.014}_{-0.016}$, and the distribution is well approximated by a Gaussian centred at 1.404 with width 0.016.}
\end{itemize}

\citet{gavazzi} find that there are two families of solutions for the lens model that lie in a degeneracy space between $\eta$, $\theta_\mathrm{s1}^{\mathrm{E}}$, and $z_\mathrm{s2}$, although \citet{sonnenfeld} break this degeneracy by modelling the stellar velocity dispersion profile of the primary lensing galaxy. However, the lensing data alone break this degeneracy if the full surface brightness distributions of the sources are reconstructed: if we start our model at the `Family II' posterior position from \citet{gavazzi} we find that the higher-redshift source is not well-focussed, but subsequently optimising the model leads to our optimal solution which agrees well (in terms of the parameters $\eta$ and $\theta_\mathrm{s1}^{\mathrm{E}}$) with the lensing-and-dynamics model of \citet{sonnenfeld}.

Both of these models find a nearly-isothermal mass distribution that is in tension with the power-law slope of $\eta = 1.196$ inferred by \citet{vegetti} using only the first source. Once again, we have started our sampling from their best-fit model and find that this model does not focus the second source; the posterior again converges to the solution presented in Table 1 and Figure 4. The reason for the discrepancy is unclear, but we note that otherwise both models are quite similar, including the magnitude and orientation of the external shear, the near-circularity of the mass density profile, and the reconstructed surface brightness distribution of the lower-redshift source; further investigation comparing our code with the code used in \citet{vegetti} is currently ongoing.

\section{Cosmological parameter constraints}
\label{sec:cosmology}

\begin{figure}
\includegraphics[trim=0.5cm 0.5cm 1cm 0.5cm, clip=true, width=0.95\columnwidth]{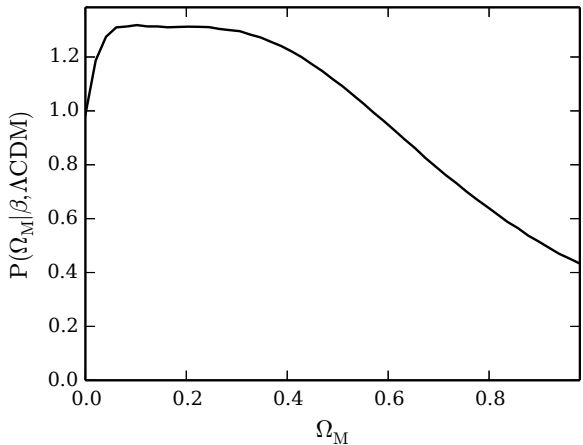}

\caption{The probability distribution function for $\Omega_{\mathrm{M}}$ given our measurement of $\beta$, marginalized over the photometric redshift for the second source. A uniform prior has been assumed for the range $0<\Omega_{\mathrm{M}}<1$.}
\label{fig:Omegam}
\end{figure}

\begin{figure}
\includegraphics[trim=0.5cm 0.5cm 0.5cm 0.5cm, clip=true, width=0.95\columnwidth]{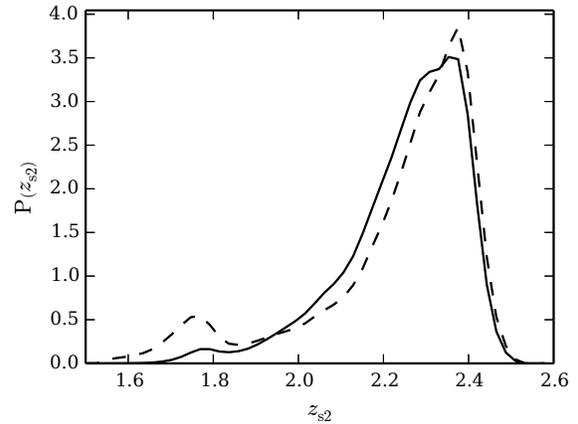}
\caption{The probability distribution function for the second source in SDSSJ0946+1006, $z_{\mathrm{s2}}$. The prior is shown dashed and is taken from the photometric redshift estimate of \citet{sonnenfeld}. The posterior is shown solid derived using our measurement of $\beta$ and assuming $\Lambda$CDM with a uniform prior on $\Omega_{\mathrm{M}}$.}
\label{fig:zs2}
\end{figure}

\begin{figure*}
\includegraphics[width=0.95\columnwidth]{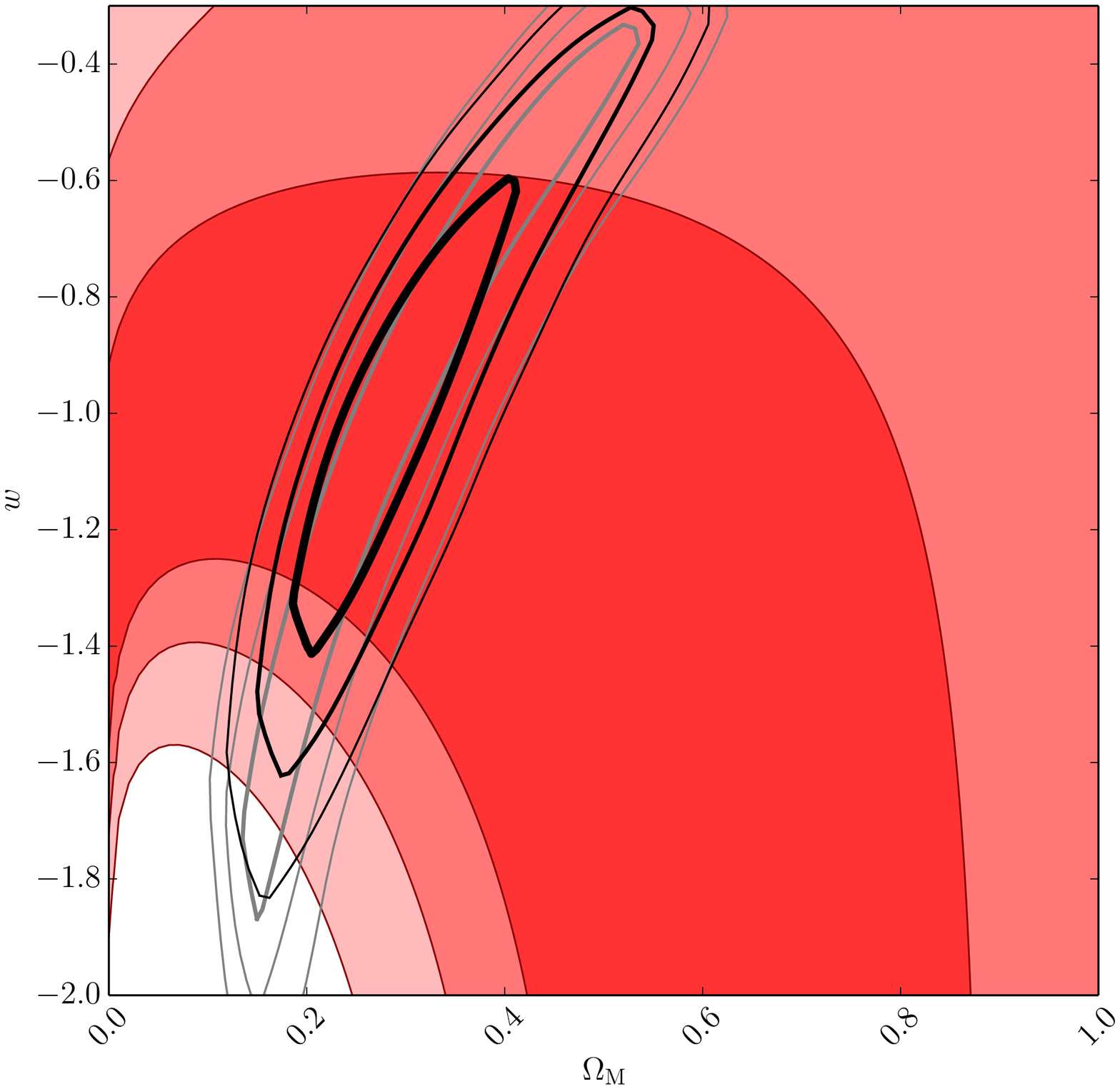}
\includegraphics[width=0.95\columnwidth]{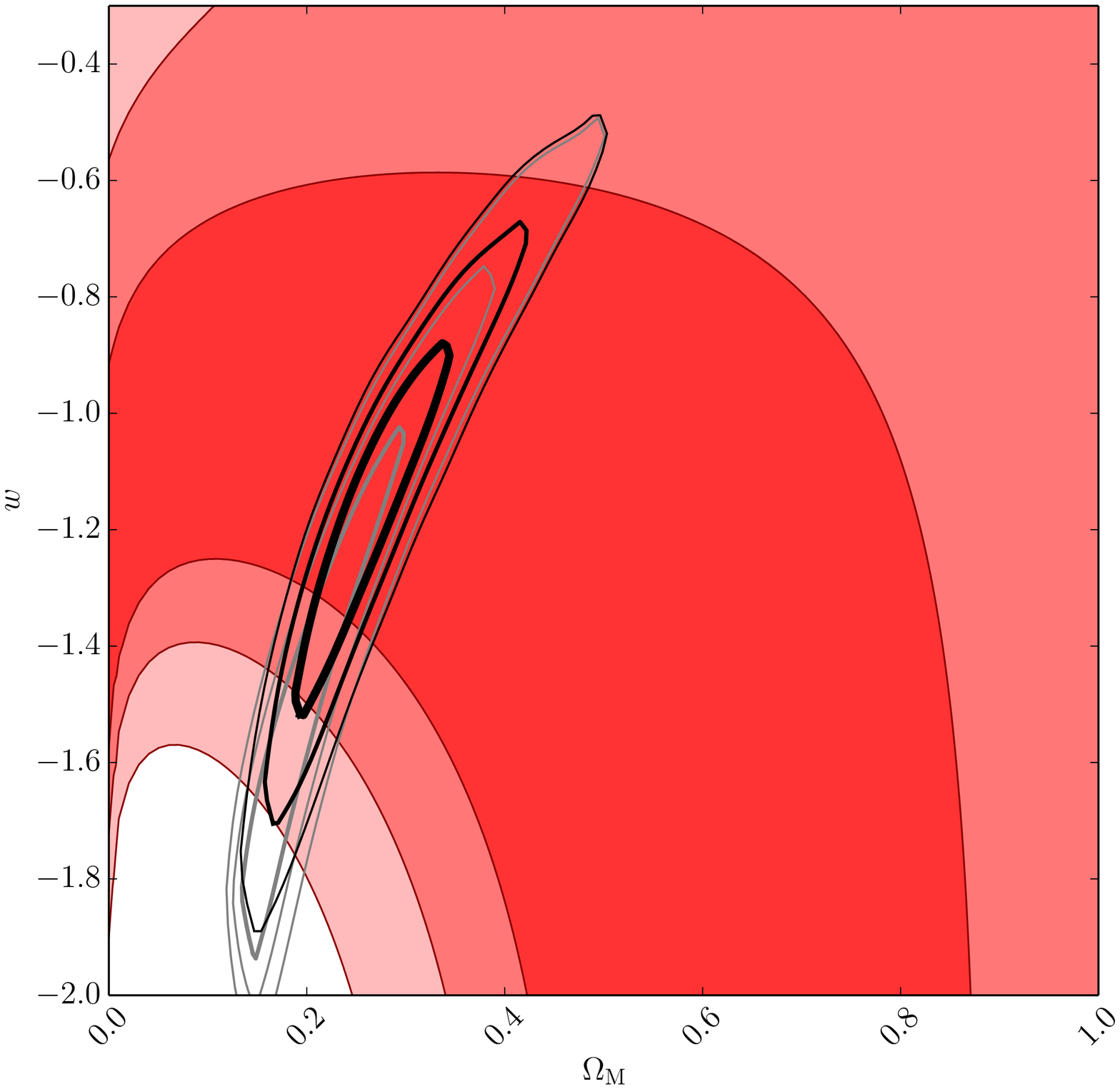}
\caption{The $w$ and $\Omega_{\text{M}}$ plane. Red shows the 68, 95 and 99.7 per cent confidence constraints derived from our measurement of the cosmological scale factor in J0946. In the left, panel grey shows the WMAP9 constraints whilst the Planck 2013 constraints are shown on the right. In both panels, black shows the combined constraint from J0946 and the CMB prior. }
\label{fig:wcdm}
\end{figure*}

Because it is a ratio of four angular diameter distances, $\beta$ is a function of only the matter content of the Universe, spatial curvature, the equation of state of dark energy, and the redshifts of the lensing and source galaxies. The lens and first source redshifts for J0946 are known from the SDSS spectroscopy \citep{gavazzi}, and \citet{sonnenfeld} used 5-band $HST$ imaging to determine a photometric redshift for the second source of $z_{s2}=2.41_{-0.21}^{+0.04}$ (68\% CL, Figure \ref{fig:zs2}). Assuming a flat $\Lambda$CDM cosmology, $\beta$ only depends on these redshifts and $\Omega_{\rm M}$, which we infer to be $\Omega_{\mathrm{M}} = 0.33_{-0.26}^{+0.33}$, with the full PDF shown in Figure \ref{fig:Omegam}.

As pointed out in \citet{collett2012}, the main benefit of using DSPLs to constrain cosmological parameters is that they have novel parameter degeneracies compared to other cosmological probes; by combining our DSPL with a CMB prior we make significantly more precise inference than with either probe individually. Adding the measurement of $\beta$ is particularly powerful for constraining non-$\Lambda$ cosmological models, where the equation of state, $w$, is not fixed to $w = -1$. Our constraints on the $w$CDM model (assuming a constant equation of state for dark energy and a flat cosmology) are shown in Figure \ref{fig:wcdm}. With a uniform prior on $w$ and $\Omega_{\mathrm{M}}$, J0946 alone is only particularly powerful at ruling out cosmologies with both low $\Omega_{\mathrm{M}}$ and very negative $w$; however this is part of the region favoured by the CMB. J0946 plus a Planck prior \citep[where we have importance sampled the constraints derived from the Planck low-l, high-l CMB temperature measurement, WMAP polarization measurement and included CMB lensing;][]{planck16} gives $w = -1.17^{+0.20}_{-0.21}$ when marginalizing over $\Omega_\mathrm{M}$, and we note that the Planck prior alone ($w = -1.49^{+0.36}_{-0.27}$) has a 50\% larger uncertainty if J0946 is not included. The J0946 constraint plus a WMAP9 prior \citep{wmap9} gives $w = -0.99^{+0.27}_{-0.25}$ compared to $w = -0.98^{+0.44}_{-0.54}$ for WMAP only.

In principle $\beta$ can be used to constrain evolving models of the dark energy equation of state, but we leave this to later work since a sample of several DSPLs is required to give interesting constraints on these models \citep{collett2012}. \new{Furthermore, we have not investigated non-flat cosmologies, since at fixed $w=-1, \Omega_\mathrm{M}=0.3, z_{\mathrm{l}}=0.2, z_{\mathrm{s1}}=0.6, z_{\mathrm{s2}}=2.3$,
\be
\frac{\del \beta}{\del \Omega_{k}}\sim 0.02{\beta}
\ee
 and $\Omega_{k}$  is already constrained at the percent level \citep[e.g][]{planck16}.
}

\section{Discussion \& Conclusion}
\label{sec:discussion}

We have derived the first cosmological constraints from a galaxy-scale double-source-plane lens. Our measurement of $\beta$ is completely independent of other cosmological probes, and can hence easily be combined with other datasets to produce tighter cosmological parameter estimates, lift parameter degeneracies, and test for the presence of unknown systematics. Because of the complementarity of DSPLs with the CMB, our measurement with just a single DSPL improves the precision of the inference on $w$ by approximately one third. More precise inferences have been made by combining the CMB with, e.g., baryon acoustic oscillation (BAO) measurements (combining Planck with the BAO results from \citealt{percival} yields $w = -1.12^{+0.10}_{-0.11}$), but we note that the number of DSPLs that will be useful for cosmological inference will increase by orders of magnitude with Euclid \citetext{Collett et al., in prep}, dramatically improving the precision but also helping to uncover systematic biases. For example, combining Planck with \textit{either} J0946 or BAO measurements causes the inference on $w$ to shift closer to $-1$ by around $1 \sigma$.

The lens model that we have presented is robustly constrained, but our inference on $\beta$ depends on the assumption that the observed lensing is entirely due to an elliptical power-law mass distribution at $z = 0.222$ and an isothermal mass distribution at $z = 0.609$. The latter point is not a significant concern here, since the highest redshift source has an impact parameter with respect to the $z = 0.609$ source that is more than three times larger than the Einstein radius $\theta^{\rm E}_{\rm s1}$. Similarly, the power-law description for the central total mass density distribution is motivated by the absence of any correlation between the power-law indices and radii of strong lenses \citep[e.g.,][]{koopmans2006,koopmans2009,auger2010}, as well as the power-law behaviour of the total mass distribution over a large range of scales from the ensemble weak lensing mass profile of lenses \citep{gavazzi2007} and mass profiles of massive X-ray-bright galaxies \citep{humphrey}. Furthermore, \citet{suyu2013} have explicitly modelled one time delay gravitational lens, RX J1131-1231, with both a power-law and a stars-plus-dark matter model and find no significant difference in the cosmographic inference between the two models when stellar kinematics are included in the modelling; \new{a similar analysis for J0946 is under way.}

\new{The mass-sheet degeneracy \citep{falco1985} will also impact our inference on $\beta$. Although multiple-source-plane lenses largely break this degeneracy for true mass sheets, Liesenborgs et al. (2008) have shown that a ring of mass (in addition to the mass from the power-law model) between the Einstein radii of the two sources can mimic the mass-sheet degeneracy even for multiple-source-plane lenses. However, it is not clear what physical process would be responsible for significant ring-like projected over- (or under-) densities and we therefore neglect this possibility.} Lensing by line-of-sight structures is also not included in our model, and if these objects introduce a positive external convergence then our estimate of $\beta$ will be low. \citet{treu2009} found no evidence that this line of sight is atypical, but even in the absence of a bias the unmodelled external convergence will lead to an artificially low uncertainty. We estimate that ignoring the external convergence results in a $\sim 1$ per cent systematic uncertainty on $\beta$ (i.e., comparable to the statistical uncertainty) which would degrade the precision of our inference on $w$ by $\approx 25$ per cent. However, directly modelling the line of sight using the existing SDSS and $HST$ imaging \citep[e.g.,][]{wong2012,collett2013} and including the velocity dispersion profile from \citet{sonnenfeld} will significantly decrease this systematic uncertainty. Furthermore, modelling the strong lensing with all of the available $HST$ data will reduce our statistical uncertainty while allowing us to further test for residual systematics by comparing our inference on $\beta$ between the different $HST$ filters.

Although there is still room for improvement of our measurement of $\beta$ for J0946, the most significant obstacle for DSPL cosmological constraints is the scarcity of simple multiple-source-plane lenses. \citet{gavazzi} suggest that one in $40-80$ galaxy-scale strong lenses should be a DSPL, and tentative Euclid forecasts of $\sim 100 000$ galaxy-galaxy strong lenses\footnote{E.g., http://euclidfrance.sciencesconf.org/file/21610} imply $\sim 2000$ DSPLs and $\sim 40$ \textit{triple}-source systems. Although it is not clear how many of these systems will be useful for cosmography, including favourable (and well-measured) lens and source redshifts \citep[e.g.,][]{collett2012}, our analysis of J0946 demonstrates the significant degeneracy-breaking power of even a single DSPL.


\section*{Acknowledgements}
 
We are greatful to the referee, Dr Prasanjit Saha, for comments on the original manuscript. TEC thanks Vasily Belokurov for supervision, guidance and suggestions. TEC acknowledges support from STFC in the form of a research studentship. MWA acknowledges support from the STFC in the form of an Ernest Rutherford Fellowship.

\label{lastpage}
\bsp

\end{document}